\documentclass[final,times,11p,twocolumn]{elsarticle}
\usepackage{amssymb}
\usepackage{amsmath}
\usepackage{lineno}
\usepackage{listings}

\journal{SoftwareX}

\begin{document}

\begin{frontmatter}

\title{OptFROG -- Analytic signal spectrograms with optimized time-frequency resolution}

\author[label1]{O. Melchert}
\ead{oliver.melchert@hot.uni-hannover.de}

\author[label1]{B. Roth}

\author[label2]{U. Morgner}

\author[label1,label2]{A. Demircan}

\address[label1]{Hannover Centre for Optical Technologies (HOT), 
                    30167 Hannover, Germany}

\address[label2]{Institute of Quantum Optics (IQO), 
                    Leibniz Universit\"at Hannover, 30167 Hannover, Germany}

\begin{abstract}
A Python package for the calculation of spectrograms with optimized time and
frequency resolution for application in the analysis of numerical simulations
on ultrashort pulse propagation is presented. Gabor's uncertainty principle
prevents both resolutions from being optimal simultaneously for a given window
function employed in the underlying short-time Fourier analysis.  
Our aim is to yield a time-frequency representation of the input signal with
marginals that represent the original intensities per unit time and frequency
similarly well. 
As use-case we demonstrate the implemented functionality for the analysis of
simulations on ultrashort pulse propagation in a nonlinear waveguide.
\end{abstract}

\begin{keyword}
Spectrogram \sep Short-time Fourier analysis \sep Analytic signal \sep Optics \sep Ultrashort pulse propagation


\end{keyword}

\end{frontmatter}


\section{Motivation and significance}
\label{sect:motivation}
The spectrogram, providing a particular time-frequency representation of
signals that vary in time \cite{Cohen:1989}, represents an inevitable tool in
the analysis of the characteristics of ultrashort optical pulses.  E.g.,
allowing to monitor the change in frequency of pulse features as function of
time permits to determine quantities that cannot be obtained from either the
time or frequency domain representation of the optical pulse alone.  The
applicability of the spectrogram to both, data retrieved from experiments
\cite{Trebino:1993, Linden:2000, Trebino:2000,Efimov:2005} where it is referred
to as frequency resolved optical gating (FROG) analysis, as well as from
numerical simulations \cite{Dudley:2002,Skryabin:2005,Guo:2013}, carried out to
complement experiments and to provide a basis for the interpretation of the
observed effects, highlights the relevance of signal processing in the field of
nonlinear optics and demonstrates the need to be able to compute such
spectrograms in the first place. Here, we consider the issue of obtaining
\emph{optimal} time-frequency representations of signals for the interpretation
of numerical experiments on ultrashort pulse propagation in nonlinear
waveguides.

In principle, a spectrogram measures the properties of the signal under
scrutiny as well as those of a user-specified window function for
localizing parts of the signal during analysis.  Exhibiting features of both,
the interpretation of the spectrogram is strongly affected by the particular
function used for windowing. Different window functions estimate different
signal properties, e.g., if a given function achieves a good approximation of
the intensities per unit time of the underlying signal, its approximation of
the intensities per unit frequency might be bad.  Consequently, the spectrogram
might suffer from distortion yielding an unreasonable characterization of the
time-frequency features of the signal under scrutiny.  The usual approach for
deciding on a particular window function is by trial-and-error and guided by
the liking and experience of the individual.

Here we present a software tool that aims at minimizing the mismatch between
the intensities per unit time and frequency and their corresponding estimates
based on the spectrogram itself, obtained for a user-supplied parameterized
window function. The resulting spectrograms are ``optimal'' in the sense that
their visual inspection exhibits a minimal amount of distortion and thus allow
for a reliable interpretation of the time-frequency composition of the input
signal. Such an approach was previously shown to result in a reasonable
characterization of the underlying time-frequency features \cite{Cohen:2003}.
It is further independent of the experience of the individual user and thus
yields reproducible results.

\section{Software description}
\label{sec:DESCRIPTION}

The presented package facilitates the construction of spectrograms for the
analytic signal (AS) \cite{Gabor:1947} $\mathcal{E}(t)$ of the real field
$E(t)$. In the Fourier domain, the angular frequency components of both are
related via $\hat{\mathcal{E}}(\omega) =  [1 +
\mathrm{sgn}(\omega)]\,\hat{E}(\omega)$ \cite{Marple:1999}.  Due to its
one-sided spectral definition the time-domain representation of the AS is
complex, further satisfying $E(t) = {\rm Re}[\mathcal{E}(t)]$.  The
construction of an AS spectrogram relies on the repeated calculation of the
spectrum of the modified signal $\mathcal{E}(t)h(t-\tau)$ at different delay
times $\tau$ in terms of the short-time Fourier transform
\begin{align}
S_\tau(\omega) = \frac{1}{\sqrt{2 \pi}} \int \mathcal{E}(t)h(t-\tau) \exp\{-i \omega t\}~{\rm d}t, \label{eq:STFT}
\end{align}
wherein $h(t)$ specifies a narrow window function centered at $t=0$ and
decaying to zero for increasing $|t|$.  The latter allows to selectively filter
parts of the AS and to estimate its local frequency content.  Scanning over a
range of delay times then yields the spectrogram as
$P_S(\tau,\omega)=|S_\tau(\omega)|^2$, providing a joint time-frequency
distribution of both, the AS and the window function \cite{WignerNote}.  For assessing the 
approximation quality of $P_S$ we utilize its time and frequency marginals
\begin{align}
&\mathsf{P}_1(\tau) = \int P_S(\tau,\omega)~{\rm d}\omega,\quad\text{and}\\
&\mathsf{P}_2(\omega) = \int P_S(\tau,\omega)~{\rm d}\tau.
\end{align}
Note that in the limit where $h(t)$ approaches a delta function, the time
marginal will approach the intensity per unit time $|\mathcal{E}(t)|^2$ but the
frequency marginal will represent the intensity per unit frequency
$|\hat{\mathcal{E}}(\omega)|^2$ only poorly. As a result, time resolution will be good
and frequency resolution will be bad, see the discussion in section \ref{sect:USECASE} below. The time-frequency uncertainty principle
prevents both resolutions from being optimal simultaneously \cite{Cohen:1989}. 

The aim of the presented package is to obtain a time-frequency representation
of the input signal for which the integrated absolute error (IAE) between its
normalized marginals and the original intensities per unit time and frequency
are minimal.  We consider a single parameter window function
$h(t,\sigma)$, e.g.\ a Gaussian function with mean $t$ and root-mean-square (rms)
width $\sigma$, and solve for 
\begin{align}
\begin{aligned}
& \sigma^\star = \underset{\sigma}{\text{arg min}}
& & Q(\sigma,\alpha) \equiv (1+\alpha) {\rm IAE}_1 + (1-\alpha) {\rm IAE}_2 \\
& \text{where}
& & {\rm IAE}_1 \equiv  \int \left| |\mathcal{E}(\tau)|^2 - \frac{\mathsf{P}_1^{(\sigma)}(\tau)}{E_S}\right|{\rm d}\tau,\\ 
& & & {\rm IAE}_2 \equiv \int \left| |\hat{\mathcal{E}}(\omega)|^2-\frac{\mathsf{P}_2^{(\sigma)}(\omega)}{E_S}\right|{\rm d}\omega.
\end{aligned}
\end{align}
Above, the underlying spectrogram is computed via $h(t,\sigma)$, indicated by
the superscript $\sigma$ on the marginals, and we assume normalization to $\int
|\mathcal{E}(t)|^2~{\rm d}t = 1$ and a total signal energy $E_S=\iint
P_S(\tau, \omega)~{\rm d}\tau\,{\rm d}\omega$ in terms of the spectrogram. 
For a good agreement of the marginals and the original intensities, the objective
function $Q$ assumes a small value.
The additional parameter $\alpha$ might be adjusted to give more weight to
frequency resolution ($\alpha<0$) or time resolution ($\alpha>0$) if
appropriate.  The particular choice $\alpha=0$ yields a balanced time-frequency
representation, see the example provided in section \ref{sect:USECASE}. The
optimized spectrogram is then computed by using $h(t)\equiv
h(t,\sigma^\star)$ for windowing.

\subsection{Software Architecture}
\label{sect:ARCHITECTURE}

OptFROG, following the naming convention \cite{PEP8} for Python packages
implemented as {\tt optfrog}, uses the Python programming language
\cite{Rossum:1995} and depends on the functionality of {\tt numpy} and {\tt
scipy} \cite{Scipy}.  It further follows a procedural programming paradigm.

\subsection{Software Functionalities}
\label{sect:FUNCTIONS}

The current version of {\tt optfrog} comprises five software units having the 
subsequent responsibilities:
\begin{description}

\item[vanillaFrog] 
    Compute a standard spectrogram $P_S(\tau,\omega)$ for the normalized 
    time-domain analytic signal for a particular window function $h(t,\sigma)$.
        
\item[optFrog] 
    Compute a time-frequency resolution optimized spectrogram for the
    normalized time-domain analytic signal using the window function
    $h(t,\sigma^\star)$ that minimizes the total IAE of both marginals. 
    Note: for the minimization of the scalar function $Q(\sigma,\alpha)$ in 
    the variable $\sigma$, the {\tt scipy} native function 
    {\tt scipy.optimize.minimize\_scalar} is employed in {\tt bounded} mode.

\item[timeMarginal]
    Compute the marginal distribution in time $\mathsf{P_1}$ based on the
    spectrogram.

\item[frequencyMarginal]
    Compute the marginal distribution in frequency $\mathsf{P_2}$ based on the
    spectrogram.

\item[totalEnergy]
    Compute the total energy $E_S$ provided by the spectrogram approximation of the
    time-frequency characteristics of the signal.
\end{description}

For a more detailed description of function parameters and return values we
refer to the documentation provided within the code \cite{optfrog_GitHub:2018}.

\begin{figure}[t!]
\centering
\includegraphics[width=\linewidth]{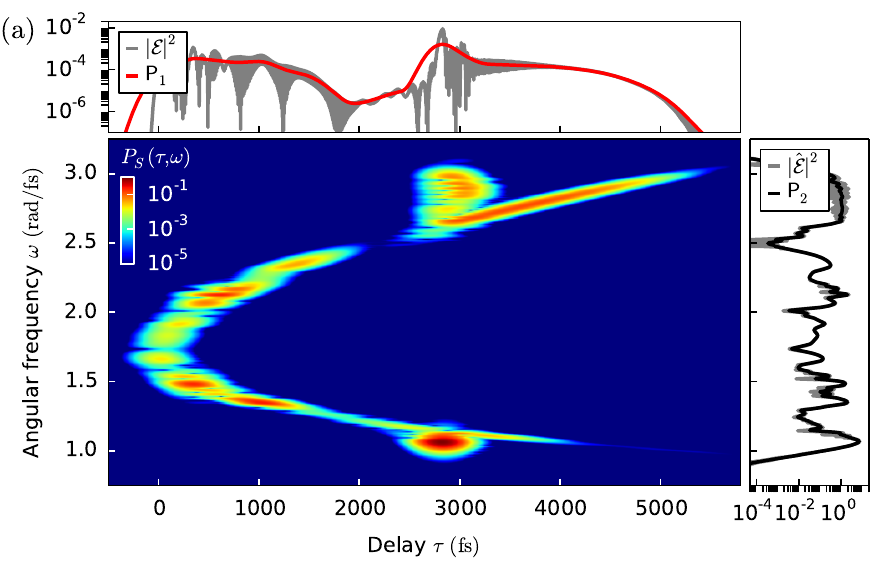}
\includegraphics[width=\linewidth]{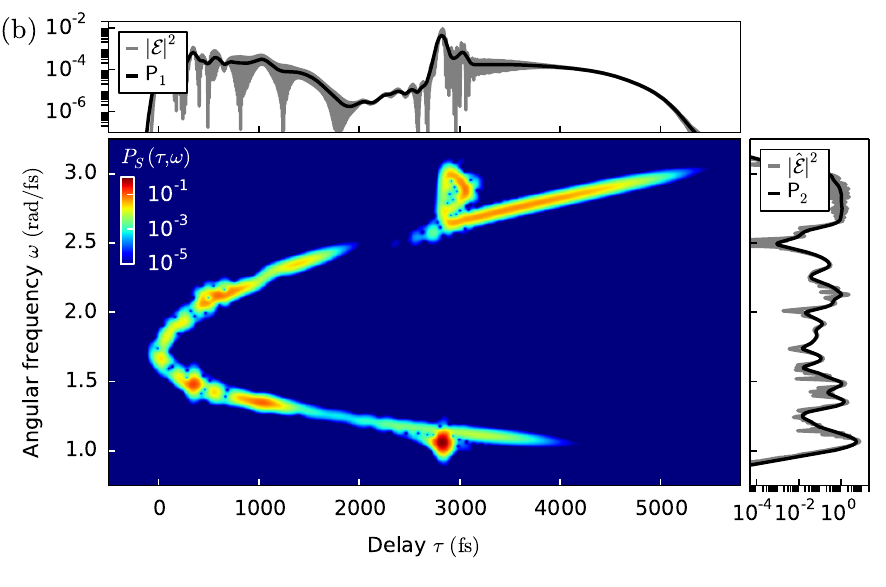}
\includegraphics[width=\linewidth]{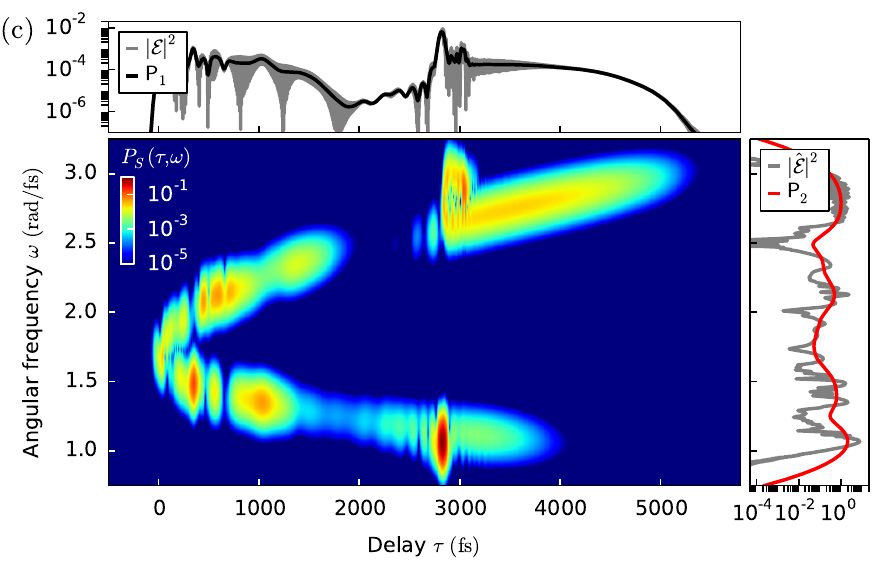}
\caption{Analytic signal spectrograms allowing for the time-frequency characterization
of a real optical field obtained from the numerical propagation of 
an ultrashort pulse in an $\mathsf{ESM}$ photonic crystal fiber.
(a) {\tt vanillaFrog}-trace for a Gaussian window function with rms-width $\sigma=140\,\mathrm{fs}$,
(b) balanced {\tt optFrog}-trace for $\sigma^\star=39.1\,\mathrm{fs}$, and, (c) 
{\tt vanillaFrog}-trace for a Gaussian window function at $\sigma=10\,\mathrm{fs}$.}
\label{fig:USECASE}
\end{figure}

\subsection{Sample code snippet}
\label{sect:CODESNIPPET}

In our research work we use {\tt optfrog} mainly in script mode.  An exemplary
data postprocessing script, reproducing Fig.\ \ref{fig:USECASE}(b) discussed in
section \ref{sect:USECASE} below, is shown in listing \ref{code:listing01}.
Therein, after importing the functionality of {\tt numpy}, {\tt optfrog}, and a
custom figure generating routine in lines 1--3, the location of the input data
(line 5) and filter options for the spectrogram output-data (lines 6p) are
specified. Note that the user defined window function (lines 9p) does not need
to be normalized. After loading the input data (lines 12p) the routine {\tt
optFrog} is used to compute an optimized spectrogram in line 15. Finally, a
visual account of the latter is prepared by the routine {\tt spectrogramFigure}
in line 17.

\begin{lstlisting}[floatplacement=H, numbers=left,
captionpos=t,keywordstyle=\bf, frame=lines,stepnumber=2,numbers=left, numbersep=5pt, xleftmargin=\parindent,
language=python,basicstyle=\ttfamily\scriptsize, 
caption={Exemplary Python script using {\tt optfrog} for the calculation 
of a time-frequency resolution optimized spectrogram.}, label=code:listing01]
import numpy as np
from optfrog import optFrog
from figure import spectrogramFigure

fName = './data/exampleData_pulsePropagation.npz'
tPars = (-500.0, 5800.0, 10)
wPars = (  0.75,   3.25,  3)

def wFunc(s0):
    return lambda x: np.exp(-x**2/2/s0/s0)

data = np.load(fName)
t, Et = data['t'], data['Et']

res = optFrog(t,Et,wFunc,tLim=tPars,wLim=wPars)

spectrogramFigure((t,Et),res)
\end{lstlisting}

\section{Illustrative Examples}
\label{sect:USECASE}

So as to demonstrate the functionality of {\tt optfrog} we
consider the numerical propagation of a short and intense few-cycle optical
pulse in presence of the refractive index profile of an ``endlessly single
mode'' ($\mathsf{ESM}$) photonic crystal fiber \cite{Birks:1997,Stone:2007}.
The underlying unidirectional propagation model includes the Kerr effect and a
delayed Raman response of Hollenbeck-Cantrell type \cite{Hollenbeck:2002}.
For the preparation of the initial condition we considered a single soliton
with duration $t_{\rm 0}=7\,\mathrm{fs}$, i.e.\ approximately $3.8$ cycles, and
soliton order $N_{\rm{s}}=8$, prepared at a center frequency $\omega=1.7\,\mathrm{rad/fs}$.
See Refs.~\cite{Amiranashvili:2010b,Amiranashvili:2011} for a detailed account
of the propagation model and Ref.~\cite{Melchert:2018} for a more thorough 
discussion of the particular problem setup. 
In Fig.~\ref{fig:USECASE} we illustrate the time-frequency characteristics of
the pulse at propagation distance $z=0.12\,\mathrm{m}$ by using a Gaussian
window function $h(t,\sigma)$ centered at $t=0$ and having rms-width $\sigma$.
Note that the delay time $\tau$ has to be interpreted as being relative to the
origin of a co-moving frame of reference in which the soliton is initially at
rest. 

In Figs.~\ref{fig:USECASE}(a,c) we demonstrate an inevitable drawback of a
trial-and-error choice of a window function used for calculating a spectrogram.
As discussed earlier, the properties of the window implies a trade-off in
resolution that might be achieved.  
I.e., if the user opts for a window function that is either too wide or too
narrow in comparison to the signal features in the time domain, only one
marginal will approximate its underlying original intensity well and, as a
result, the spectrogram will appear distorted.  This is shown in
Fig.~\ref{fig:USECASE}(a), where a {\tt vanillaFrog} trace using
$\sigma=140\,\mathrm{fs}$ demonstrates a good frequency resolution and a bad
time resolution. Conversely, as evident from Fig.~\ref{fig:USECASE}(c), a {\tt
vanillaFrog} trace using $\sigma=10\,\mathrm{fs}$ yields a good time resolution
and a bad frequency resolution.
In contrast, if the IAEs of both marginals are minimized simultaneously by
aid of a numerical algorithm, both marginals of the optimized spectrogram are
found to approximate the original intensities per unit time and frequency
similarly well. Consequently, the resulting spectrogram provides a most
reasonable time-frequency representation of the underlying signal. To
demonstrate this, the balanced ($\alpha=0$) {\tt optFrog} trace for the 
optimized window function, obtained for $\sigma^\star=39.1\,\mathrm{fs}$ with
$Q=0.39$, is shown in Fig.~\ref{fig:USECASE}(b).

\section{Impact}
\label{sect:IMPACT}

Computing reliable spectrograms represents an integral part in the analysis of
the characteristics of ultrashort optical pulses.  The publicly available and
free Python package {\tt optfrog} performs the nontrivial task of computing
such spectrograms with optimized time-frequency resolution.  It is based on a
computational approach to parameter optimization in opposition to common
trial-and-error approaches, helping to save time and effort and yielding
reproducible results independent of the skill of the individual user.
It addresses researchers in the field of ultrashort pulse propagation and related 
disciplines where signal analysis in terms of short-time Fourier transforms is 
of relevance. 
As independent software postprocessing tool it is ideally suited
for the analysis of output data obtained by existing pulse propagation codes,
as, e.g., the open source  {\tt LaserFOAM} (Python) \cite{Amorim:2009} and {\tt
gnlse} (Matlab) \cite{Travers:2010} solver for the generalized nonlinear
Schr\"odinger equation. 

\section{Conclusions}
\label{sect:conclusions}

The {\tt optfrog} Python package provides easy-to-use tools that yield a
time-frequency representation of a real valued input signal and allow to
quantify how well the resulting spectrogram approximates the signal under
scrutiny for a user supplied window function.  

We have shown how {\tt optfrog} can be used to calculate analytic signal based
spectrograms that are optimal in the sense that their visual inspection
exhibits a minimal amount of distortion, allowing for a reliable interpretation
of the time-frequency composition of the input signal.  

The {\tt optfrog} software tool, including scripts that implement the exemplary
use-cases illustrated in section \ref{sect:USECASE}, is available for download
and installation under Ref.\ \cite{optfrog_GitHub:2018}. 

\section*{Acknowledgements}

We acknowledge support from the VolkswagenStiftung within the
Nieders\"achsisches Vorab program (HYMNOS; Grant ZN 3061) and 
Deutsche Forschungsgemeinschaft (Grand MO 850/20-1). 

\section*{References}

\bibliography{myReferences} 
\bibliographystyle{unsrt}  





\clearpage
\section*{Required Metadata}
\label{}

\section*{Current code version}
\label{}

Ancillary data table required for subversion of the codebase. Kindly replace examples in right column with the correct information about your current code, and leave the left column as it is.

\begin{table}[!h]
\begin{tabular}{|l|p{6.5cm}|p{6.5cm}|}
\hline
\textbf{Nr.} & \textbf{Code metadata description} & \textbf{Please fill in this column} \\
\hline
C1 & Current code version & 1.0.0 \\
\hline
C2 & Permanent link to code/repository used for this code version & https://github.com/omelchert/optfrog.git \\
\hline
C3 & Legal Code License   & MIT License \\
\hline
C4 & Code versioning system used & none \\
\hline
C5 & Software code languages, tools, and services used & Python, GitHub  \\
\hline
C6 & Compilation requirements, operating environments \& dependencies & The {\tt optfrog} package requires Python, numpy and scipy. The installation process requires Pythons setuptool package and the provided use-cases need Pythons matplotlib for figure generation.\\
\hline
C7 & If available Link to developer documentation/manual & Documentation provided within code \\
\hline
C8 & Support email for questions & oliver.melchert@hot.uni-hannover.de\\
\hline
\end{tabular}
\caption{Code metadata (mandatory)}
\label{} 
\end{table}

\end{document}